\documentclass[12pt]{article}
\usepackage{graphicx}
\usepackage{enumerate}
\def\beq{\begin{equation}}
\def\eeq#1{\label{#1}\end{equation}}
\def\eeqn{\end{equation}}

\def\beqa{\begin{eqnarray}}
\def\eeqa#1{\label{#1}\end{eqnarray}}
\def\eeqan{\end{eqnarray}}

\let\bar=\overbar

\def\Dslash{\not{\hbox{\kern-4pt $D$}}}
\def\dslash{\not{\hbox{\kern-2pt $\del$}}}

\def\msb{{\bar{\ssstyle M \kern -1pt S}}}

\def\Title#1{\begin{center} {\Large {\bf #1} } \end{center}}
\begin{document}
\noindent
\footnotesize
{Proceedings of CKM 2012, the 7th International Workshop on the CKM
Unitarity Triangle,\\
University of Cincinnati, USA, 28 September - 2 October 2012}
\normalsize

\bigskip\bigskip
\Title{$V_{ud}$ from Nuclear Mirror Transitions}
\bigskip


\begin{center}
{\large Oscar Naviliat-Cuncic\index{Naviliat-Cuncic, O.}}\\
{\small\it NSCL and Department of Physics and Astronomy,
MSU, East Lansing, 48824 MI, USA}
\end{center}

\bigskip
\begin{abstract}
We review the determination of $V_{ud}$ from nuclear mirror
transitions and describe current experimental and theoretical efforts aimed at
improving its precision.
\end{abstract}

\section{Introduction}
The test of the Conservation of the Vector Current (CVC)
in nuclear $\beta$ decay and its relation to the determination
of the Cabibbo
angle has been the focus of much experimental and theoretical
activity during more than 50 years \cite{Tow10}. Precision measurements in
super-allowed pure Fermi transitions and refined theoretical
corrections have enabled to extract the value of $V_{ud}$ with a
precision of $2\times10^{-4}$ \cite{Tow10}.

In addition to Fermi transitions, super-allowed transitions
between isospin doublets
provide another set of nuclear transitions to test the CVC hypothesis
and to extract
$V_{ud}$ \cite{Nav09}. Such transitions occur between mirror
nuclei and are
driven by the vector ($V$) and axial-vector ($A$) interactions.

This contribution reviews the determination of $V_{ud}$
from nuclear mirror transitions and describes current experimental
and theoretical efforts aimed at improving its precision.

\section{Mirror transitions}
\begin{figure}[!hbt]
\centerline{
\includegraphics[scale=0.65]{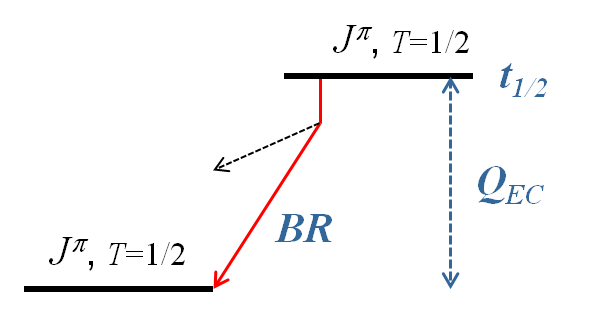}
}
\caption{Scheme of a mirror transition indicating the spectroscopic
quantities that are required to determine $V_{ud}$ (see text
for details).} 
\label{fig:mirror-transition}
\end{figure}

$\beta$-decay transitions between isobaric analogue states within an
isospin doublet are called ``nuclear mirror transitions''. 
The initial and final states have therefore
the same spins and parities (Fig.\ref{fig:mirror-transition}). 
The extent to which the $V$ and $A$ components contribute to the transition
is described by the mixing
ratio, $\rho = C_A M_{GT}/(C_V M_F)$, where $M_{GT}$
and $M_F$ are respectively the Gamow-Teller (GT) and
Fermi nuclear matrix elements and $C_A$ and $C_V$ are the axial and vector
effective couplings respectively.

Except for the neutron and
$^3$H, all mirror transitions proceed by $\beta^+$ emission. The number of
protons and neutrons in the parent nucleus are related by $Z=N+1$ so that
the nuclei of interest lie on the proton rich side of the nuclear chart.


The master formula that relates $V_{ud}$ to the corrected
decay rate in a nuclear transition can be casted in the
form \cite{Nav09}
\begin{equation}
{\cal F}t = \frac{K}{G^2_F g^2_V V^2_{ud} ( 1 + \Delta_R^V )} ,
\label{eq:Ft}
\end{equation}
where
$K$ is a constant, $G_F$ is the Fermi constant, 
and $\Delta_R^V$ is a transition-independent radiative correction
and whose values can be found elsewhere \cite{Nav09}.
The interest in writing the relation between 
${\cal F}t$ and $V_{ud}$ following Eq.(\ref{eq:Ft}) is that,
if CVC holds, then $g_V = 1$ and
the right-hand side is independent of the decay.
This is valid for pure Fermi transitions, for nuclear mirror
transitions and also for neutron decay.
The left-hand side of Eq.(\ref{eq:Ft}) depends, in turn, on the
decaying system. For a nuclear mirror transition it
takes the form \cite{Nav09}

\begin{equation}
{\cal F}t^m = f_V t (1+\delta^\prime_R)(1+\delta_{\rm NS}-\delta_C)
\left[ 1 + (f_A/f_V)\rho^2 \right]
\label{eqn:Ft-mirror}
\end{equation}
where $f_V$ is the statistical rate function for the vector part
\cite{Har05},
$t$ is the partial half-life of the transition, $\delta^\prime_R$ and
$\delta_{\rm NS}$ are transition-dependent contributions to the
radiative corrections,
$\delta_C$ is the isospin symmetry breaking (ISB) correction,
and $f_A/f_V$ is the ratio between the
statistical rate
functions for the axial and the vector parts \cite{Sev08}.

The spectroscopic quantities required to determine
${\cal F}t^m$ are shown in Fig.\ref{fig:mirror-transition}.
The $Q_{EC}$-values determine the
statistical rate functions $f_V$ and $f_A$. The branching ratio,
$BR$, and half-life, $t_{1/2}$ determine the partial half-life, $t$.

\section{Experimental status}
\begin{figure}[!htb]
\centerline{
\includegraphics[width=0.75\textwidth]{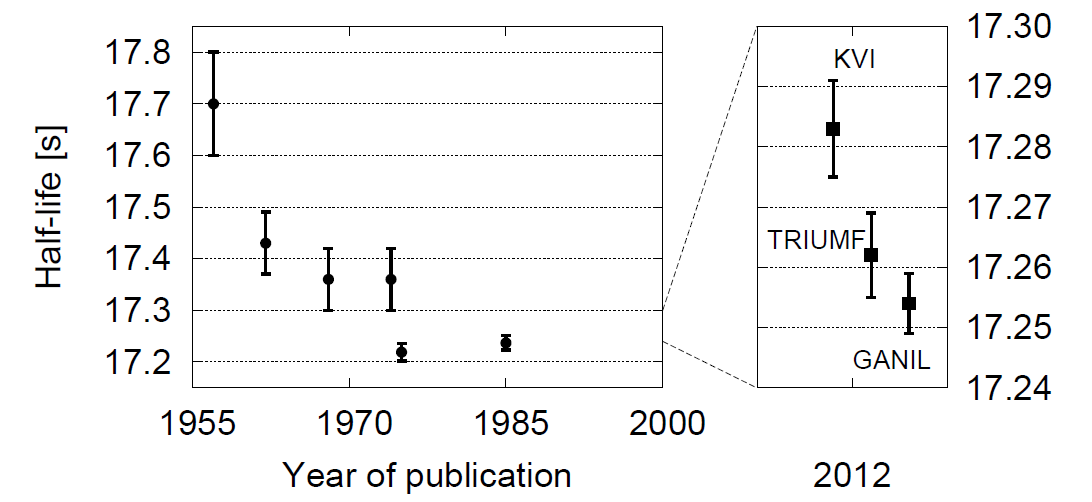}
}
\caption{Evolution of the values of the $^{19}$Ne half-life (see text for details).}
\label{fig:Ne19-halfLife}
\end{figure}

The possibility offered by mirror transitions to determine
$V_{ud}$ has motivated new measurements of
the spectroscopic quantities.
An example of recent progress, which confirms the potential
offered by nuclei for high precision experiments, is provided
by measurements
of the $^{19}$Ne half-life. Figure \ref{fig:Ne19-halfLife}
shows the values of the half-life adopted in the survey of
Ref.~\cite{Sev08} as a function of the year of publication.
The zoom for year 2012 shows new results from measurements carried
out at
KVI \cite{Bro12}, TRIUMF \cite{Tri12} and GANIL \cite{Uji12},
with relative uncertainties smaller than $5\times 10^{-4}$.
The technique used at GANIL, based on the implantation of nuclei
in a superconductor, offers additional room
for further improvements \cite{Uji12}.

Preliminary measurements of the $^{21}$Na and $^{37}$K lifetimes have
been attempted at KVI but no detailed analysis has yet been
completed \cite{Bro12}. A  new measurement of the $^{39}$Ca lifetime 
has been carried out \cite{Bla10}; the result did not significantly improve
the precision on the world average value.
Measurements of the mass \cite{Kan10} and of the lifetime \cite{Bac12}
of $^{31}$S have been performed at Jyv\"{a}skyl\"{a} providing new
data for the determinations of the $Q_{EC}$-value and the partial
half-live.

Although the error on the current value of $V_{ud}$ obtained from mirror
transitions is dominated by the error of the mixing
ratio \cite{Sev11}, it would be extremely
useful to improve previous mass values deduced from reaction experiments,
by using Penning trap mass spectrometry, in order to dispose of a robust
data set.

The mixing ratio can be extracted from measurements of correlation coefficients
like e.g.\ the $\beta$-$\nu$ angular correlation, $a$,
the $\beta$ asymmetry parameter, $A$, or the $\nu$ asymmetry
parameter.

The group from LPC in Caen has initiated a dedicated program at GANIL
to measure $a$ in several nuclei.
The setup uses a transparent Paul trap surrounded by detectors to record
the recoil ions and the $\beta$ particles \cite{Cou12}.
A measurement of $a$ in $^{35}$Ar decay
has recently been completed. The collected data corresponds to a statistical
precision of about 0.4-0.5\%. The group is currently preparing
a measurement of $a$ in $^{19}$Ne decay with the same setup.

Experiments with polarized nuclei can access the parameter $A$.
For several mirror candidates (e.g. $^{21}$Na, $^{23}$Mg, $^{29}$P,
$^{35}$Ar, $^{37}$K),
the mixed transition between the isobaric analogue states
is accompanied by a pure GT transition to an excited state in the
daughter nucleus, with a branching ratio of few \%.
Some of these nuclei can be polarized by
optical pumping and be implanted in suitable targets. The measurement
of $A$ for $\beta$ particles detected in coincidence with the
de-exciting $\gamma$ ray provides a measurement of the initial nuclear
polarization. Such relative measurements of $A$ will become
feasible at the new BECOLA (BEam COoler and LAser spectroscopy) end-station
at NSCL \cite{Min12} and will provide new and improved determinations
of $\rho$.

\section{Theoretical corrections}
\begin{figure}[!htb]
\centerline{
\includegraphics[width=0.85\textwidth]{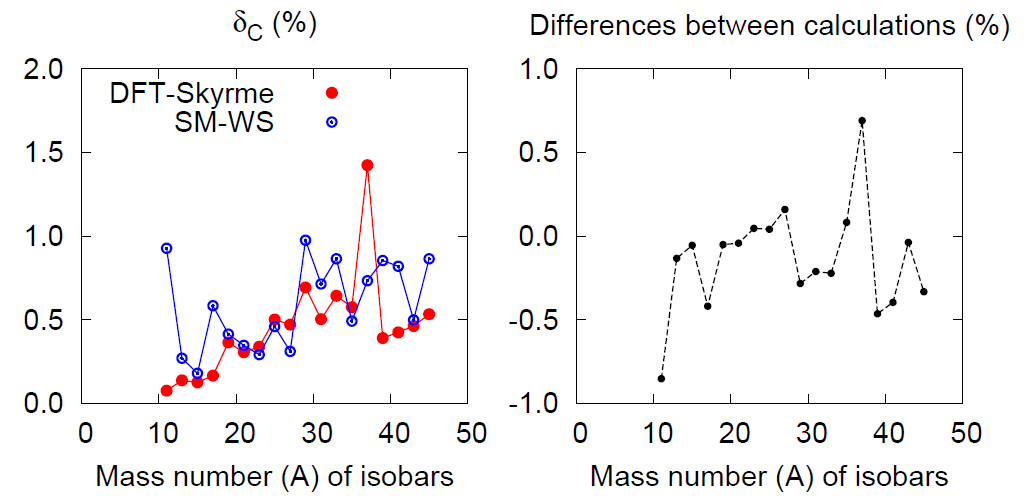}
}
\caption{Calculated ISB corrections (see text for details).}
\label{fig:theoCorr}
\end{figure}

The corrected decay rates, Eq.(\ref{eqn:Ft-mirror}), require the
inclusion of small
nuclear structure and ISB
corrections.
The first systematic determination of such corrections for mirror nuclei
used shell-model
Wood-Saxon wave functions \cite{Sev08}. The ISB corrections have recently been
calculated for both pure Fermi and mirror transitions, using density functional
theory with independent Skyrme interaction \cite{Sat12}. The results
of the two calculations are shown in Fig.~\ref{fig:theoCorr} (left panel)
along with the difference between the two calculations (right panel).

Among the five parent nuclei ($^{19}$Ne, $^{21}$Na, $^{29}$P, $^{35}$Ar,
$^{37}$K)
considered so far to determine $V_{ud}$ from mirror
transitions \cite{Nav09}, the largest difference
between the two calculations arises for $^{37}$K and mounts to 0.7\%.
This has however a small impact on the average value, $\bar{{\cal F}t}^m$,
and
hence on $V_{ud}$ since the data from this transition has currently
a smaller weight due to the precision of its mixing ratio.

\section{Summary and Outlook}
The value of $V_{ud}$ deduced from mirror transitions is \cite{Nav09}

\begin{equation}
V_{ud} = 0.9719(17)
\label{eqn:Vud}
\end{equation}
where the precision is dominated by the experimental error on the mixing
ratios.

Two new results of the $^{19}$Ne lifetime have recently been published
\cite{Tri12,Uji12} and new systematic calculations of ISB corrections have
been performed \cite{Sat12}.  Measurements of spectroscopic quantities
are required to build up a robust data set while theoretical
ISB and nuclear structure corrections are crucial to estimate the
associated theoretical uncertainties.

Progress towards a more accurate determination of $V_{ud}$ from mirror
transitions definitely relies on improved correlation measurements for
the extraction of $\rho$.
Current measurements of the $\beta\nu$ angular correlation in a Paul trap
and future relative $\beta$-asymmetries measurements with optically pumped
low energy beams will provide significant contributions to this end.


\end{document}